\documentclass[aps,prd,twocolumn,groupedaddress,nofootinbib]{revtex4-1}

\usepackage{dcolumn}

\usepackage{amssymb}

\usepackage{amsmath}

\usepackage{bm}

\usepackage{hyperref}

\usepackage{color}

\usepackage{graphicx}

\definecolor{MyBlue}{rgb}{0.15,0.15,0.70}
\hypersetup{
colorlinks=true,
citecolor=MyBlue,
linkcolor=MyBlue,
urlcolor=MyBlue
}

\begin{document}

\title{Observed Angles and Geodesic Light-Cone Coordinates}

\author{Ermis Mitsou} 
\email{ermitsou@physik.uzh.ch}
\author{Fulvio Scaccabarozzi}
\email{fulvio@physik.uzh.ch}
\author{Giuseppe Fanizza} 
\email{gfanizza@physik.uzh.ch}

\affiliation{Center for Theoretical Astrophysics and Cosmology,
Institute for Computational Science \\ University of Z\"urich, Winterthurerstrasse 190, CH-8057, Z\"urich, Switzerland}

\begin{abstract}
We discuss the interpretation of the angles in the Geodesic Light-Cone (GLC) coordinates. In particular, we clarify the way in which these angles can be identified with the observed ones.
We show that, although this identification is always possible in principle, one cannot implement it in the usual gauge-fixing way, i.e. through a set of conditions on the GLC metric. Rather, one needs to invoke a tetrad at the observer and a Cartesian-like coordinate system in order to obtain the desired map globally on the observed sky. 
\end{abstract}

\maketitle

\section{Introduction}
Most of the physical information relevant for cosmology is carried by photons emitted by distant sources. 
By measuring the properties of these photons, we discover the structure of the Universe, its composition and the history of its evolution, with implications that go backwards to its origin. 
However, in order to interpret correctly and optimize this information, it is essential to understand how the photons' propagation is affected by the inhomogeneities in the Universe.  
Many efforts are made towards this direction, both in theory and observations. 

In order to simplify the theoretical description of light propagation, the Geodesic Light-Cone (GLC) coordinates were introduced in \cite{GLC2011}. 
The GLC coordinates are indeed adapted to describe the observation of light sources lying on the past light-cone of an observer. Their remarkable properties allow to describe analytically the photons' propagation accounting for inhomogeneities. 
Thanks to this, the GLC coordinates have been used to obtain the expressions of light-cone observables and to perform light-cone average, showing great advantage in the derivation \cite{GLC0,GLC1,GLC2,GLC3,GLC4,GLC5,GLC6,GLC7,GLC8,Fleury:2016}.

Despite the fact that many possible applications of GLC coordinates have been successfully explored, there is still room for extensions or further study of the coordinates themselves.  
It has been recently noted that the angles of GLC coordinates are \textit{not in general} (and not by construction) the observed angles \cite{Fulvio}, contrary to what was commonly believed. 
A simple way to see this is that the observed angles are defined up to a rotation of the observed sky, while the GLC angles are defined up to a full reparametrization. 
 However, the validity of the results obtained in the GLC coordinates still relies on the fact that, by fixing the gauge freedom of GLC coordinates, it is possible to identify the GLC angles with the observed ones, i.e. the so-called ``observational gauge'' \cite{GLC4,Fleury:2016}. 
 
One would therefore expect that there exists a gauge condition {\it on the metric components}, i.e. on top of the GLC conditions, with which the GLC angles are the ones the actual observer uses.  In particular such a gauge condition should reduce the residual freedom of angular reparametrizations to the rotation group.
 As we shall see, the problem is very subtle and involves a careful geometrical analysis. 
Our conclusion is that, although the observational gauge exists, it cannot be expressed as a further gauge condition of the GLC metric. Rather, we will see that one needs to go beyond the GLC framework, i.e invoke extra structure such as another coordinate system and/or a tetrad at the observer, in order to find the relation between the GLC angles and the observed ones (see \cite{Fulvio}).

The structure of the paper is the following. In sec.~\ref{sec:GLC}, we review the GLC coordinates, describing their properties and features. In sec.~\ref{sec. obs angle}, we define rigorously the observed angles as those measured by the observer in the orthonormal rest frame. In sec.~\ref{sec: GLC vs OBS}, through a purely geometrical argument, we show that the observational gauge cannot be expressed as conditions on the metric components. We conclude with a discussion in sec.~\ref{sec: disc}. Finally, appendix \ref{A} contains the detailed derivation of the GLC residual gauge symmetries. 

\section{The geodesic light-cone gauge}  \label{sec:GLC}

The GLC coordinates $x^\mu=(w,\tau,\theta^a)$, $a \in \{ 1,2 \}$, are defined by the line element
\begin{equation} \label{eq:GLC}
ds^2 = \Upsilon^2 dw^2 -2 \Upsilon dw  d\tau + \gamma_{ab} (d\theta^a-U^a dw)(d\theta^b-U^b dw) \, ,
\end{equation}
and the requirement that the topology of the $w,\tau =$ constant hypersurfaces $S_{w,\tau}$ is spherical. The $\theta^a$ are therefore angular coordinates and $\gamma_{ab}(w,\tau,\theta^a)$ is a two-metric on $S_{w,\tau}$. It is then convenient to define the vector fields
\begin{equation} \label{eq:GLCframe}
u^\mu := -g^{\mu\nu}\partial_\nu \tau  \, , \qquad k^\mu:=g^{\mu\nu}\partial_\nu w \, ,
\end{equation}
which are normal to the $\tau, w = {\rm const}.$ hypersurfaces, respectively. The fact that
\begin{equation}
g(u,u) \equiv -1 \, , \qquad g(k,k) \equiv 0 \, , 
\end{equation}
implies that the $\tau$ parameter foliates space-time into space-like hypersurfaces, while the $w$ parameter foliates space-time into \textit{light-like cones}, since the $\mathbb{S}_{w,\tau}$ must be spheroidal. One important property of the GLC coordinates is that the $u^{\mu}$ and $k^{\mu}$ vectors are also geodesic, by construction\footnote{For a derivation of the GLC line element from these geometrical properties, see \cite{Fleury:2016}.},
\begin{equation}
u^{\nu} \nabla_{\nu} u^{\mu} \equiv 0 \, , \qquad k^{\nu} \nabla_{\nu} k^{\mu} \equiv 0 \, .
\end{equation}
This means that $\tau$ is the proper-time of a free-falling observer family with four-velocity $u^\mu$, while $w$ actually foliates space-time into \textit{light-cones}, because of the spherical topology of the $S_{w,\tau}$ surfaces. In particular, $k^{\mu}$ is then proportional to the momentum vectors of photons traveling towards the cone tips. In order to have a unique set of coordinates associated to each space-time point, we need to choose either past or future light-cones so, in GLC, the former is chosen. 

The other important property of the GLC coordinates is that the aforementioned light-like geodesics, on top of having a constant $w$ along their path
\begin{equation} \label{eq:consttheta}
k^{\mu} \partial_{\mu} w \equiv 0 \, ,
\end{equation}
also have constant angular coordinates
\begin{equation} \label{eq:consttheta}
k^{\mu} \partial_{\mu} \theta^a \equiv 0 \, .
\end{equation}
Now note that this non-trivial foliation of space-time implies non-trivial ranges for the $\tau, w$ coordinates. Given a specific past light-cone $w$, we get that $\tau$ is defined on a semi-infinite interval $\tau \in (-\infty, T(w)]$, for some value $T(w)$ corresponding to the tip of the cone $w$. One can therefore think of $T(w) - \tau$ as a ``radius" on the past light-cone $w$. Alternatively, for a given hypersurface $\tau$, the $w$ parameter has also semi-infinite range $[ W(\tau),+\infty)$, where $W(\tau)$ is the inverse map of $T(w)$, i.e. $W(T(w)) \equiv w$. One can thus think of $w - W(\tau)$ as the ``radius" on the $\tau = {\rm const}.$ hypersurfaces. 

We then have that the spheroids $S_{w,\tau}$ collapse to points at the tips of the past light-cones, so the angular coordinates $\theta^a$ are not defined there. All fields must therefore be angle-independent at the tip points and the latter are uniquely determined by the pair of coordinates $(w, T(w))$, or alternatively $(W(\tau),\tau)$. By construction, these points form a time-like geodesic ${\cal G}$ with proper time $\tau$ and four-velocity $u^{\mu}(\tau) := u^{\mu}(W(\tau),\tau)$, i.e. the world-line of some specific free-falling observer. For later arguments it will be convenient to pick a specific observer point $o$ on that world-line and denote by $\tau_o$ and $w_o \equiv W(\tau_o)$ the corresponding coordinates. 

Now the singular behavior of the $S_{w,\tau}$ submanifolds imposes regularity conditions on the metric at ${\cal G}$ if we want the geometry to be smooth there. To find them, we note that in Cartesian-like coordinates all metric components are dimensionless so, in spherical-like coordinates such as the GLC ones, some of the metric components acquire length dimensions to compensate the fact that the angular coordinates are dimensionless. Close to ${\cal G}$, these length dimensions are given by the ``radii" discussed previously. Thus, regularity implies that a metric component with length dimension $L$ behaves like $\sim [w - W(\tau)]^L$ when $\tau$ is held constant, and $\sim [T(w) - \tau]^L$ when $w$ is held constant, as we approach ${\cal G}$. In particular, defining $U_a := \gamma_{ab} U^b$, we have that 
\begin{equation} \label{eq:gamma}
\left\{ U_a, \gamma_{ab}, \partial_{\tau} \gamma_{ab}, \partial_w \gamma_{ab} \right\}({\cal G}) \equiv 0 \, .
\end{equation}
To understand this intuitively, note the analogy with the behavior of the angular two-metric in the case of Euclidean space, which goes like $\sim r^2$.

Finally, note that eq. (\ref{eq:GLC}) does not completely determine the coordinate system, i.e. one can still perform a subset of coordinate transformations that do not change the form of the GLC line-element. These are commonly referred to as ``residual gauge symmetries" and consist of the most general coordinate transformations that preserve the form of the GLC metric \textit{and} do not depend on the metric components themselves, so that they hold for arbitrary space-time. We find\footnote{For a more detailed discussion about the GLC residual gauge freedom, see \cite{GLC4,Fleury:2016}.} (see appendix \ref{A})
\begin{equation}\label{gauge freedom}
\begin{split}
& \tau \rightarrow \tau + \text{const.} \,,
\\
& w \rightarrow w'(w) \,,  
\\
&\theta^a \rightarrow \theta'^a(w,\theta) \,,
\end{split}
\end{equation}  
under which the metric components transform as
\begin{eqnarray} 
\Upsilon & \to & \frac{\partial w}{\partial w'}\, \Upsilon \, , \label{eq:Upsilontrans} \\
U^a & \to & \frac{\partial w'}{\partial w} \left[ \frac{\partial \theta'^a}{\partial \theta^b}\, U^b - \frac{\partial \theta'^a}{\partial w} \right] \, , \\
\gamma^{ab} & \to & \frac{\partial \theta'^a}{\partial \theta^c} \frac{\partial \theta'^b}{\partial \theta^d}\,\gamma^{cd} \, ,   \label{eq:gammatrans}
\end{eqnarray}  
in agreement with \cite{Fleury:2016}.

\section{Observed angles}\label{sec. obs angle}

In this section we consider a generic coordinate system, but use again the symbol $k^{\mu}$ to denote the four-momentum of photons. It is a well-known fact that the observer performs measurements with respect to an orthonormal frame at $o$, i.e. a ``tetrad''. We will denote these four vectors by $e_A^{\mu}$, where $A \in \{ 0,1,2,3 \}$ indexes the basis elements, and the orthonormality condition then reads
\begin{equation} \label{eq:tetrad}
g_o(e_A, e_B) \equiv g_{\mu\nu,o} \, e_A^{\mu} e_B^{\nu} = \eta_{AB} \, .
\end{equation}
On top of this condition, we also have that the time-like element $e_0^{\mu}$ must coincide with the four-velocity of the observer $u_o^{\mu}$ so that the spatial elements $e_I^{\mu}$, where $I \in \{ 1,2,3 \}$, span the rest-frame of the observer. Note that the role of the tetrad can be played by the coordinate vectors $\partial_{\mu}$ at $o$ if the coordinates are such that the metric is Minkowksi there, since the scalar product is then $g_o(\partial_{\mu}, \partial_{\nu}) \equiv g_{o,\mu\nu} = \eta_{\mu\nu}$. Put differently, in such a system the tetrad matrix is the identity matrix. This is for instance the case of Fermi coordinates around the observer's geodesic.

Now what the observer actually measures is the four-momentum of incoming photons at $o$ in the tetrad basis, i.e. $k_o^A := e^A_{\mu} k_o^{\mu}$, where $e_{\mu}^A$ is the inverse matrix of $e_A^{\mu}$. Using the fact that $k^{\mu}$ is light-like we get
\begin{equation}
\eta_{AB} k_o^A k_o^B \equiv 0 \, ,
\end{equation}
and this allows us to express it in terms of observed frequency $\omega_o$ and angles $\theta^a_o \in \{\theta_o, \varphi_o \}$
\begin{equation}
k_o^A \equiv \omega_o ( 1, - n^I_o ) \, ,
\end{equation}
where
\begin{equation}
n^I_o := ( \sin \theta_o \cos \varphi_o, \sin \theta_o \sin \varphi_o, \cos \theta_o ) \, .
\end{equation}
The $\theta^a_o$ angles therefore parametrize the unit-normed elements of $T_o {\cal M}$ that are normal to $e_0^{\mu} \equiv u_o^{\mu}$, i.e. in the observer's rest-frame. The corresponding subset of the vector space $T_o {\cal M}$ is therefore the unit sphere $\mathbb{S}_o$ and corresponds to the mathematical definition of the {\it observed sky}. Since $n_o^I(\theta_o)$ is a map from $\mathbb{S}_o$ to 3D Euclidean space, it induces a line-element on $\mathbb{S}_o$, the trivial one
\begin{eqnarray} \label{eq:dlS}
d l^2_{\mathbb{S}_o} & := & [\delta_{IJ} \, \partial_a n^I_o \partial_b n_o^J ] \, d \theta^a_o d \theta^b_o \nonumber \\
 & = & d\theta_o^2 + \sin^2\theta_o \, d\varphi_o^2 \, .
\end{eqnarray}
This is indeed the metric that the actual observer implicitly uses when performing measurements: when giving the angular separation between two sources, when measuring a solid angle or when integrating to get spherical harmonics components. Finally, note that the tetrad vectors are determined up to rotation of the spatial elements
\begin{equation}
e_I^{\mu} \to R_I^{\,\,J} e_J^{\mu} \, , \qquad \Rightarrow \qquad n^I_o \to R^I_{\,\,J} n^J_o \, ,
\end{equation}
since we must preserve $e_0^{\mu} \equiv u_o^{\mu}$, which corresponds to the SO(3) freedom the actual observer has in setting up the $\theta^a_o$ parameters on the sky. Thus, $\mathbb{S}_o$ is a manifold that only admits coordinate systems $(\theta_o, \varphi_o)$ related by global rotations.

\section{Observed vs. GLC angles}\label{sec: GLC vs OBS}

Let us now consider the bundle of light-like geodesics $\gamma$ reaching the observer $o$, which is therefore a topological space homeomorphic to a sphere, and we denote it by $B_o$. In the GLC coordinates, we have that
\begin{equation} 
\dot{\gamma}^{\mu}(\lambda) \sim k^{\mu}(\gamma(\lambda)) \, , 
\end{equation}
for some affine parameter $\lambda$. Since $k^{\mu} =  - \Upsilon^{-1} \delta^\mu_\tau$, we have that these paths have constant $w, \theta^a$ parameters
\begin{equation} 
\gamma^w(\lambda) = w_o \, , \hspace{1cm} \gamma^a(\lambda) = \theta^a \, , \hspace{1cm} \forall \lambda \, ,
\end{equation}
as already discussed in section \ref{sec:GLC}. In particular, the fact that the GLC angles $\theta^a$ are constant along each $\gamma$ path means that, even though they are space-time coordinates, they can {\it also} serve as a parametrization of $B_o$, i.e. to every element $\gamma \in B_o$ one associates a unique $\theta^a(\gamma)$. 

Now the elements of $B_o$ are in a one-to-one correspondence with the elements of $\mathbb{S}_o$, i.e. every light-like geodesic path reaching $o$ corresponds to a point on the observed sky and vice-versa. In fact, $B_o$ serves as the usual definition of the observer sky in the absence of the observer tetrad \cite{Low, Perlick, KK}. The advantage of the tetrad construction $\mathbb{S}_o$ is that it also provides the parametrization and metric the actual observer uses on that manifold. With the $\mathbb{S}_o \sim B_o$ identification, the observed angles $\theta_o^a$ can also serve as an alternative parametrization $\theta_o^a(\gamma)$ of $B_o$. The relation between the two parametrizations is then a coordinate transformation $\theta^a \to \theta^a_o(\theta)$ on $B_o$. Now remember that the GLC angles are defined up to an arbitrary reparametrization $\theta^a \rightarrow \theta'^a(w_o,\theta)$ (see eq. \eqref{gauge freedom}). This means that, by performing a residual gauge transformation, the $\theta^a_o(\theta)$ relation can actually be {\it any} bijective map. In particular, one can then use this freedom to set
\begin{equation} 
\theta^a(\gamma) = \theta_o^a(\gamma) \, ,
\end{equation}
i.e. to turn the GLC angles into the angles the actual observer uses to parametrize the sky. This is known as the ``observational gauge" \cite{Fleury:2016}. The problem with this manipulation, however, is that one needs to know the map $\theta^a_o(\gamma)$ beforehand, i.e. one needs to consider the tetrad construction described in the previous section to make the connection \cite{Fulvio}. The question we wish to answer here is whether it is possible to determine the $\theta^a_o(\gamma)$ map {\it within the} GLC {\it framework alone}, i.e. whether there exists a residual gauge condition on the GLC metric that makes the GLC angles observed angles. 

A first possibility for relating GLC angles to the observed ones would be to perform the tetrad construction at $o$ in GLC coordinates. The problem with this approach, however, is that the GLC metric is singular at $o$, because there $\gamma_{ab} = 0$, so we cannot define the tetrad matrix $e^A_{\mu}$. On top of this, we have that the photon four-momentum does not carry angular information in GLC coordinates since $k^{\mu} = - \Upsilon \delta^{\mu}_{\tau}$. Rather, this angular information now lies in the coordinates of the geodesic $\gamma^a$.

The second possibility is to work with the notion of observed sky instead. In the general tetrad approach $\mathbb{S}_o$ was defined as the set of unit-normed elements of $T_o {\cal M}$ that are normal to $u_o^{\mu}$. In particular, this means that one can picture $\mathbb{S}_o$ as an infinitesimal 2D submanifold ${\cal S}_o \subset {\cal M}$ around $o$ with the following properties:
\begin{itemize}
\item
At any point $P \in {\cal S}_o$, the tangent space $T_P {\cal S}_o$ is normal to $u^{\mu}(P)$.

\item
${\cal S}_o$ has constant intrinsic curvature.
\end{itemize}
The analogous object in the GLC formalism is the closest spheroid surface to the observer $o$ on the observed light-cone $w_o$, i.e. $S_o := S_{w_o,\tau_o - d \tau}$, whose line-element is
\begin{equation} \label{eq:gammao}
dl^2_{S_o} = \gamma_{ab}(w_o,\tau_o - d \tau,\theta) \, d \theta^a d \theta^b \, .
\end{equation}
If $S_o$ can be given the geometric properties described above, then we would have found a way to match the GLC and observed angles. More precisely, if $S_o$ has constant curvature, then there exists an angular reparametrization which leads to a trivial line-element, i.e. one can impose the gauge condition 
\begin{equation}  \label{eq:gaugecond}
dl^2_{S_o} \sim d \theta^2 + \sin^2 \theta\, d \varphi^2 \, ,
\end{equation}
thus making the GLC angles observed angles. 

For the first condition, we note that the vectors $\partial_a$ generate, by definition, the tangent spaces of the $w,\tau = {\rm const.}$ hypersurfaces that are the $S_{w,\tau}$ spheroids. We then have the identity
\begin{equation} 
g(u, \partial_a) \equiv 0 \, ,
\end{equation}
all over space-time, so we can consistently extend this result to the infinitesimal spheroid $S_o$. Thus, the first criterion is automatically satisfied, i.e. the GLC angles are measures in the observer's rest-frame.

We next consider the second condition, so we define the intrinsic curvatures of the $S_{w,\tau}$ spheroids, i.e. the 2D Ricci scalars ${}^{(2)} R_{w,\tau}(\theta)$ of $\gamma_{ab}(w,\tau,\theta)$, which contain the full curvature information since ${}^{(2)} R_{abcd}[\gamma] \equiv \gamma_{a[c}\gamma_{d]b} {}^{(2)} R[\gamma]$ in two dimensions. Now the fact that ${}^{(2)} R_{w,\tau}$ scales as the inverse area of $S_{w,\tau}$ implies that its limit at $o$ is singular. A well-defined expression for the condition of constant curvature at the observer is therefore 
\begin{equation}  \label{eq:Rcond}
\lim_{\tau \to \tau_o}[\partial_a {}^{(2)} R]_{w_o,\tau}(\theta) = 0 \, , \qquad \forall \theta \, ,
\end{equation}
which is a 2D covector equation. Note that, contrary to (\ref{eq:gamma}), this is not a regularity condition that is automatically satisfied by the requirement of a smooth space-time. Indeed, one can easily find examples of coordinate systems that foliate a smooth manifold in such a singular manner. For example, if one considers ellipsoidal coordinates $(r, \theta, \varphi)$ in Euclidean space, then the $r = {\rm const}.$ hypersurfaces have non-constant intrinsic curvature all the way down to $r = 0$.

Now note that the 2D intrinsic curvatures ${}^{(2)} R_{w,\tau}$ are not a set of scalars under generic 4D coordinate transformations, because the latter alter the foliation and thus the geometry/topology of the $w,\tau = {\rm const}.$ surfaces. The question now is whether the GLC residual gauge freedom contains such coordinate transformations that allow one to modify the intrinsic geometry of $S_o$. To answer this, we observe that the two-metric $\gamma_{ab}$ transforms as a 2D tensor under these residual gauge transformations (\ref{eq:gammatrans}). The geometry of the $S_{w,\tau}$ spheroids is therefore unchanged, and in particular the one of $S_o$. As a result, there is no way to impose the second criterion (\ref{eq:Rcond}) through a residual gauge transformation.

Note, however, that the residual gauge transformations are only a subset of the transformations that preserve the GLC form, because they must also be independent of the metric components, i.e. of the space-time under consideration. So there remains the question of whether there exists a metric-dependent transformation that preserves the GLC form and also allows us to reach the condition (\ref{eq:Rcond}). Going back to appendix \ref{A}, this means that we must impose the four conditions \eqref{cond} for the four functions \eqref{func} \textit{plus} the local condition (\ref{eq:Rcond}). Even though the latter concerns an infinitesimal space-time region, it is enough to make this PDE system over-determined. Therefore, for a \textit{generic} space-time, no solution exists.    

We conclude that, for generic space-times, the geometry of $S_o$ is in general different from the one of the observed sky $\mathbb{S}_o$. Although $S_o$ lies in the observer's rest-frame by construction, there is no residual symmetry that can give it a constant intrinsic curvature. There is therefore no residual gauge condition that can guarantee that the GLC angles $\theta^a$ correspond to observed angles. In other words, the observational gauge exists, because of the full angular reparametrization symmetry, but one cannot express it as a condition on the GLC metric components. 

In practice, when computing cosmological observables using GLC coordinates, this obstruction manifests itself as follows. First, note that the two-metric ``at" the observer (\ref{eq:gammao}), usually denoted ``$\gamma_{ab}(\lambda_o)$" in the GLC literature, appears indeed in several observables (e.g. luminosity distance, lensing shear, etc.), since the observer is one of the two boundaries of the photon path. If equation (\ref{eq:gaugecond}) could be reached by the residual gauge symmetry, then this would correspond to the gauge condition of the observational gauge, i.e. a further equation the metric components must satisfy on top of the GLC ones. The fact that we cannot impose (\ref{eq:gaugecond}) in general means that it is not straightforward to obtain the ``true" observables, i.e. those that are expressed in terms of observed angles, out of the general expressions derived in the GLC literature. One cannot simply replace ``$\gamma_{ab}(\lambda_o)$" by (\ref{eq:gaugecond}).

We finally note that the above conclusion is global in nature, i.e. we cannot obtain the desired metric on {\it all} of $S_o$. However, as with any differentiable manifold, one can perform an angular reparametrization to set normal coordinates around a definite $\theta_*^a$ point on $S_o$ 
\begin{equation} \label{eq:limgaugecond}
\gamma_{ab}(w_o,\tau_o - d \tau,\theta_*) \sim \delta_{ab} + {\cal O}[(\theta - \theta_*)^2] \, .
\end{equation}
With geodesic normal coordinates, one can even extend this condition along a full geodesic $\theta_*^a(\lambda)$ on $S_o$, but not more. One can thus work with the observed angle parametrization {\it within the GLC framework} by imposing (\ref{eq:limgaugecond}) on $\gamma_{ab}$ in the observables of interest, but {\it only} around a given source direction $\theta_*^a$, or a one-dimensional family of such directions at most. Therefore, this construction is not enough if one is interested in correlation functions of cosmological observables, which require evaluation at several arbitrary points on the observed sky.

\section{Discussion}\label{sec: disc}

Let us now draw the practical consequences for the computations within the GLC framework and, in particular, let us discuss the extra structure that is required in order to get to the observational gauge. After having performed all the desired computations, the equations are invariant under the residual gauge transformations. This means that we are working with an arbitrary angular parametrization, which is therefore not the one the observer uses in general. One must therefore use the freedom in reparametrizing the angles (\ref{gauge freedom}) to express the equations in terms of the observed angles. We are then confronted with the fact that the constant curvature condition (\ref{eq:Rcond}) cannot be reached by a residual gauge transformation. Nevertheless, we know that there exists a gauge in which $\theta^a(\gamma) = \theta_o^a(\gamma)$ since we have the full set of angular reparametrizations at our disposal. In order to find the $\theta_o^a(\gamma)$ map, we must thus go beyond the GLC framework, i.e. we need to invoke the tetrad vectors at the observer (see \cite{Fulvio}). More precisely, one first needs to consider some other ``Cartesian-like" coordinate system $x'^{\mu}$ in which the metric is non-singular at $o$ and $k'^{\mu}$ contains the angular information of the photon geodesic, in order to define a non-singular tetrad and get observed angles through $k^A_o = e'^A_{\mu} k'^{\mu}_o$. One can then finally use the relation $k'^{\mu}(\gamma, k)$ to express the observed angles in $k^A_o$ in terms of the GLC angles in $\gamma^a$. Using a coordinate system which directly satisfies $g_{o,\mu\nu} = \eta_{\mu\nu}$ kills two birds with one stone because then the tetrad matrix is the identity. In practice, however, it is usually convenient to use a FLRW-like coordinate system, in which case the tetrad is non-trivial. 

Now since the generic $\theta^a$ are related to $\theta^a_o$ by some reparametrization $\theta^a \rightarrow \theta'^a(\theta)$, this ambiguity will appear as ``integration constants'' at $o$ of the form $C^a(\theta)$ in the perturbative solutions to cosmological observables {\it in the $x'^{\mu}$} coordinates \cite{Fulvio}. These functions are thus fixed by the requirement (\ref{eq:Rcond}) and therefore depend on the gravitational fields at (and dynamics of) the observer through the tetrad. They will thus bring in terms (perturbations) evaluated at the observer position, precisely as in the geometric approach \cite{Yoo:2009,Yoo:2010,Yoo}. This argument supports the importance of considering terms at the observer, usually neglected in the literature, as they can acquire a concrete physical meaning (see \cite{Biern:1,Biern:2}), or be important in order to restore the correct gauge transformation of physical observables \cite{FBY}.

\acknowledgments

We wish to thank Sang Gyu Biern, Pierre Fleury, Fabien Nugier and Jaiyul Yoo for providing beneficial comments about this work. We are especially grateful to Maurizio Gasperini, Giovanni Marozzi and Gabriele Veneziano for detailed discussions. E.M. is supported by the Tomalla foundation, F.S. is supported by the Swiss National Science Foundation and
E.M. and G.F. are supported by a Consolidator Grant of the European Research Council (ERC-2015-CoG grant 680886).

\onecolumngrid

\appendix

\section{Residual gauge symmetries of GLC coordinates}
\label{A}

To find the GLC residual gauge symmetries, we take the most general coordinate transformations: 
\begin{equation}\label{func}
	w'=w'(w,\tau,\theta^a),
	\qquad\qquad \tau'=\tau'(w,\tau,\theta^a), \qquad\qquad \theta'^a=\theta'^a(w,\tau,\theta^a),
\end{equation}
and impose that the form of GLC metric is preserved. The metric in the new coordinates is given by
\begin{equation}\label{ds2}
	\begin{split}
		ds'\,^2 &=  \mathcal A'\, dw'\,^2 - 2\, \mathcal B'\, dw' d\tau'  - 2\,\mathcal C'_a \, dw' d\theta'^a  + \mathcal D'_{ab} \, d\theta'^a d\theta'^b
		 \\
 &= \mathcal A\, dw^2  - 2\,\mathcal B \, dw d\tau  - 2 \,\mathcal C_a \, dw d\theta^a  + \mathcal D_{ab} \, d\theta^a d\theta^b 
 + \mathcal E \,  d\tau^2 + 2\, \mathcal F_a \, d\tau d\theta^a \,, 
 	\end{split}
\end{equation}
where
\begin{equation}
	\mathcal A' :=  \Upsilon'\,^2 + \gamma'_{ab} \,U'^a U'^b \,,
	\qquad\quad
	\mathcal B' := \Upsilon'\,,
	\qquad\quad
	\mathcal C'_a := \gamma'_{ab} U'^b \,,
	\qquad\quad
	\mathcal D'_{ab} := \gamma'_{ab} \,,
\end{equation}
and
\begin{eqnarray}
	\mathcal A &:= & \mathcal A' \bigg(\frac{\partial w'}{\partial w} \bigg)^2  - 2\, \mathcal B' \,  \frac{\partial w'}{\partial w}\frac{\partial \tau'}{\partial w}  
 - 2\, \mathcal C'_a \, \frac{\partial \theta'^a}{\partial w}\frac{\partial w'}{\partial w}  + \mathcal D'_{ab} \frac{\partial \theta'^a}{\partial w}\frac{\partial \theta'^b}{\partial w} \,,
 \\
 \mathcal B &:= & - \mathcal A' \, \frac{\partial w'}{\partial w}\frac{\partial w'}{\partial \tau} + \mathcal B' \bigg( \frac{\partial w'}{\partial w}  \frac{\partial \tau'}{\partial \tau}  +  \frac{\partial w'}{\partial \tau} \frac{\partial \tau'}{\partial w} \bigg)
   + \mathcal C'_a \, \bigg( \frac{\partial \theta'^a}{\partial w} \frac{\partial w'}{\partial \tau} +  \frac{\partial \theta'^a}{\partial \tau} \frac{\partial w'}{\partial w} \bigg) -  \mathcal D'_{ab}\frac{\partial \theta'^a}{\partial w}\frac{\partial \theta'^b}{\partial \tau}  \,,
 \\
\mathcal C_a &:= &  -\mathcal A' \, \frac{\partial w'}{\partial w}\frac{\partial w'}{\partial \theta^a}   +  \mathcal B' \bigg( \frac{\partial w'}{\partial w}  \frac{\partial \tau'}{\partial \theta^a}  +   \frac{\partial w'}{\partial \theta^a}  \frac{\partial \tau'}{\partial w} \bigg)   
+  \mathcal C'_b \, \bigg(\frac{\partial \theta'^b}{\partial w}\frac{\partial w'}{\partial \theta^a}    +   \frac{\partial \theta'^b}{\partial \theta^a}\frac{\partial w'}{\partial w} \bigg) - \mathcal D'_{bc}\frac{\partial \theta'^b}{\partial w} \frac{\partial \theta'^c}{\partial \theta^a}  \,,
 \\
\mathcal D_{ab} &:= & \mathcal A' \, \frac{\partial w'}{\partial \theta^a} \frac{\partial w'}{\partial \theta^b}   - 2\, \mathcal B' \, \frac{\partial w'}{\partial \theta^{(a}}\frac{\partial \tau'}{\partial \theta^{b)}}
 -  2\, \mathcal C'_c \, \frac{\partial \theta'^c}{\partial \theta^{(a}} \frac{\partial w'}{\partial \theta^{b)}} + \mathcal D'_{cd} \frac{\partial \theta'^c}{\partial \theta^a}  \frac{\partial \theta'^d}{\partial \theta^b}  \,,
 \\
\mathcal E &:= & \mathcal A' \bigg( \frac{\partial w'}{\partial \tau} \bigg)^2  - 2\, \mathcal B' \, \frac{\partial w'}{\partial \tau}\frac{\partial \tau'}{\partial \tau}
   - 2\, \mathcal C'_a \, \frac{\partial \theta'^a}{\partial \tau} \frac{\partial w'}{\partial \tau} + \mathcal D'_{ab} \frac{\partial \theta'^a}{\partial \tau}  \frac{\partial \theta'^b}{\partial \tau}  \,,
 \\
\mathcal F_a &:= &  \mathcal A' \, \frac{\partial w'}{\partial \tau}\frac{\partial w'}{\partial \theta^a}       
 -  \mathcal B' \bigg( \frac{\partial w'}{\partial \tau}\frac{\partial \tau'}{\partial \theta^a} + \frac{\partial w'}{\partial \theta^a}\frac{\partial \tau'}{\partial \tau} \bigg)      
   -  \mathcal C'_b \, \bigg( \frac{\partial \theta'^b}{\partial \tau}\frac{\partial w'}{\partial \theta^a} 
+ \frac{\partial \theta'^b}{\partial \theta^a} \frac{\partial w'}{\partial \tau}\bigg) + \mathcal D'_{bc} \frac{\partial \theta'^b}{\partial \tau}  \frac{\partial \theta'^c}{\partial \theta^a} \,.
\end{eqnarray}
The four conditions which we need to impose to preserve the form of the metric are 
\begin{equation}\label{cond}
	 \mathcal E = 0 \,, \qquad\qquad\qquad  \mathcal F_a=0 \,, \qquad\qquad\qquad \mathcal B^2 + \mathcal C^2 = \mathcal A \,,
\end{equation}
where $\mathcal C^2 = \mathcal C \, \mathcal D^{-1}\, \mathcal C = \mathcal C_a \, \mathcal D^{ab} \, \mathcal C_b $.
Since, by definition, the solution cannot depend on the metric components, these conditions imply the following constraints for the Jacobians of the transformations
\begin{equation}
	 \frac{\partial w'}{\partial \tau} =0 \,, 
 \qquad\qquad
 \frac{\partial \theta'^a}{\partial \tau} = 0 \,,
\qquad\qquad
	 \frac{\partial w'}{\partial \theta^a} = 0 \,,
\qquad\qquad
	 \frac{\partial \tau'}{\partial w} = 0 \,,
\qquad\qquad \frac{\partial \tau'}{\partial \tau} = 1 \,.
\end{equation}
The most general solution reads
\begin{equation}
	w \rightarrow w'(w) \,, \qquad\qquad \tau \rightarrow  \tau + f(\theta)\,, \qquad\qquad \theta^a \rightarrow \theta'^a(w,\theta) \,.
\end{equation}
However, for a given value of $w$, as $\tau$ approaches the tip value $T(w)$, $f(\theta)$ must tend to a constant since angles are not defined there. Consequently, $f(\theta) \equiv \text{const.}$ and the $\tau$ coordinate is determined up to constant time translation.

\end{document}